\documentclass{emulateapj}
\usepackage{graphicx}
\usepackage{amssymb,amsmath}
\usepackage{longtable}
\usepackage{color}
\usepackage{verbatim}
\begin{document}
\title{NANOGrav Constraints on Gravitational Wave Bursts With Memory}
\author{
Z.~Arzoumanian\altaffilmark{1}, 
A.~Brazier\altaffilmark{2}, 
S.~Burke-Spolaor\altaffilmark{3,20},
S.~J.~Chamberlin\altaffilmark{4},
S.~Chatterjee\altaffilmark{2},
B.~Christy\altaffilmark{5},
J.~M.~Cordes\altaffilmark{2},
N.~J.~Cornish\altaffilmark{6}, 
P.~B.~Demorest\altaffilmark{3},
X.~Deng\altaffilmark{7},
T.~Dolch\altaffilmark{2},
J.~A.~Ellis\altaffilmark{8,21},
R.~D.~Ferdman\altaffilmark{9},
E.~Fonseca\altaffilmark{10},
N.~Garver-Daniels\altaffilmark{11},
F.~Jenet\altaffilmark{12}, 
G.~Jones\altaffilmark{13},
V.~M. Kaspi\altaffilmark{9}, 
M.~Koop\altaffilmark{7},
M.~T.~Lam\altaffilmark{2},
T.~J.~W.~Lazio\altaffilmark{8},
L.~Levin\altaffilmark{11},
A.~N.~Lommen\altaffilmark{5}, 
D.~R.~Lorimer\altaffilmark{11},
J.~Luo\altaffilmark{12},
R.~S.~Lynch\altaffilmark{14},
D.~R.~Madison\altaffilmark{2,$\dagger$},
M.~A.~McLaughlin\altaffilmark{11}, 
S.~T.~McWilliams\altaffilmark{11},
D.~J.~Nice\altaffilmark{15},
N.~Palliyaguru\altaffilmark{11},
T.~T.~Pennucci\altaffilmark{16},
S.~M.~Ransom\altaffilmark{17}, 
X.~Siemens\altaffilmark{4}, 
I.~H.~Stairs\altaffilmark{10}, 
D.~R.~Stinebring\altaffilmark{18},
K.~Stovall\altaffilmark{19},
J.~Swiggum\altaffilmark{11},
M.~Vallisneri\altaffilmark{8},
R.~van~Haasteren\altaffilmark{8,21},
Y.~Wang\altaffilmark{12}, \&
W.~W.~Zhu\altaffilmark{10}.
	}

\collaboration{NANOGrav Collaboration\altaffilmark{$\star$}}

\altaffiltext{$\star$}{Author order alphabetical by surname}

\altaffiltext{$\dagger$}{Contact author: D. R. Madison, drm252@cornell.edu}

\altaffiltext{1}{Center for Research and Exploration in Space Science and Technology and X-Ray Astrophysics Laboratory, NASA Goddard Space Flight 
Center, Code 662, Greenbelt, MD 20771, USA}

\altaffiltext{2}{Department of Astronomy, Cornell University, Ithaca, NY 14853, USA}

\altaffiltext{3}{National Radio Astronomy Observatory, 1003 Lopezville Rd., Socorro, NM 87801, USA}

\altaffiltext{4}{Center for Gravitation, Cosmology and Astrophysics, Department of Physics, University of Wisconsin-Milwaukee, P.O. Box 413, Milwaukee, WI 53201, USA}

\altaffiltext{5}{Department of Physics and Astronomy, Franklin \& Marshall College, P.O. Box 3003, Lancaster, PA 17604, USA}

\altaffiltext{6}{Department of Physics, Montana State University, Bozeman, MT 59717, USA}

\altaffiltext{7}{Department of Astronomy and Astrophysics, Pennsylvania State University, University Park, PA 16802, USA}

\altaffiltext{8}{Jet Propulsion Laboratory, California Institute of Technology, 4800 Oak Grove Drive, Pasadena, CA 91106, USA}

\altaffiltext{9}{Department of Physics, McGill University, 3600  University St., Montreal, QC H3A 2T8, Canada}

\altaffiltext{10}{Department of Physics and Astronomy, University of British Columbia, 6224 Agricultural Road, Vancouver, BC V6T 1Z1, Canada}

\altaffiltext{11}{Department of Physics and Astronomy, West Virginia University, P.O. Box 6315, Morgantown, WV 26506, USA}

\altaffiltext{12}{Center for Gravitational Wave Astronomy, University of Texas at Brownsville, Brownsville, TX 78520, USA}

\altaffiltext{13}{Department of Physics, Columbia University, New York, NY 10027, USA}

\altaffiltext{14}{National Radio Astronomy Observatory, P.O. Box 2, Green Bank, WV 24944, USA}

\altaffiltext{15}{Department of Physics, Lafayette College, Easton, PA 18042, USA}

\altaffiltext{16}{University of Virginia, Department of Astronomy, P.O. Box 400325, Charlottesville, VA 22904, USA}

\altaffiltext{17}{National Radio Astronomy Observatory, 520 Edgemont Road, Charlottesville, VA 22903, USA}

\begin{comment}
\altaffiltext{18}{Department of Physics and Astronomy, Oberlin College, Oberlin, OH 44074, USA}

\altaffiltext{20}{Department of Physics and Astronomy, University of New Mexico, Albuquerque, NM 87131, USA}

\altaffiltext{21}{Jansky Fellow}

\altaffiltext{22}{Einstein Fellow}
\end{comment}

\begin{abstract}
Among efforts to detect gravitational radiation, pulsar timing arrays are uniquely poised to detect ``memory" signatures, permanent perturbations in spacetime from highly energetic astrophysical events such as mergers of supermassive black hole binaries. The North American Nanohertz Observatory for Gravitational Waves (NANOGrav) observes dozens of the most stable millisecond pulsars using the Arecibo and Green Bank radio telescopes in an effort to study, among other things, gravitational wave memory.  We herein present the results of a search for gravitational wave bursts with memory (BWMs) using the first five years of NANOGrav observations. We develop original methods for dramatically speeding up searches for BWM signals. In the directions of the sky where our sensitivity to BWMs is best, we would detect mergers of binaries with reduced masses of $10^9$ $M_\odot$ out to distances of 30 Mpc; such massive mergers in the Virgo cluster would be marginally detectable. We find no evidence for BWMs. However, with our non-detection, we set upper limits on the rate at which BWMs of various amplitudes could have occurred during the time spanned by our data--e.g., BWMs with amplitudes greater than $10^{-13}$ must occur at a rate less than 1.5 yr$^{-1}$. 
\end{abstract}
\maketitle

\section{Introduction}

Due to the intrinsically nonlinear nature of Einstein's equations, all systems that radiate gravitational waves (GWs) are anticipated to produce ``memory," non-oscillatory components of the gravitational waveforms \citep{s77,bp79,bt87,c91,bd92}. Supermassive black hole binaries (SMBHBs), during the final few orbits preceding their merger, are expected to generate GW bursts with memory (BWMs) with sufficiently large amplitudes to make them potentially detectable with pulsar timing arrays \citep[PTAs;][]{f09,s09,pbp10,vl10,cj12,mcc14}.  \citet{cbv+14} recently singled out memory as a key detection target for PTAs as a probe of exotic and unexpected GW sources like phase transitions in the early Universe.

To facilitate GW detection, several international consortia are currently using sensitive radio telescopes paired with pulsar-optimized hardware to realize the PTA concept \citep{hd83,fb90}. The European Pulsar Timing Array \citep[EPTA;][]{kc13}, the North American Nanohertz Observatory for Gravitational Waves \citep[NANOGrav;][]{m13}, and the Parkes Pulsar Timing Array \citep[PPTA;][]{h13} are pushing precision pulsar timing to its limits and developing new data analysis techniques to usher in the era of PTA GW astronomy. By pooling their data and expertise, these consortia have formed the International Pulsar Timing Array \citep[IPTA;][]{m13_2} which is poised to become the most sensitive of all PTAs.

In recent years, the various PTAs have begun to place astrophysically meaningful upper limits on continuous GWs from individually resolvable SMBHBs \citep{abb+14,zhw+14} and a stochastic background of GWs \citep{vlj+11,dfg+13,src+13}. \citet{whc+14} have searched for BWMs in the first approximately six years of PPTA data; they detected nothing, but determined with 95\% confidence that BWMs with amplitudes greater than $10^{-13}$ occur at a rate less than 0.8 yr$^{-1}$.

In this paper, we search for BWMs in the first approximately five years of NANOGrav data using new techniques that lead to considerable computational speedups over the search methods described by \citet{mcc14} and those used by \citet{whc+14}. In Section~2, we describe the data used in our BWM search. In Section~3, we describe the BWM signal model we use for our analysis. In Section~4, we discuss models of noise and how our sensitivity to BWMs is influenced by them.  In Sections~5 and 6, we describe our search techniques, differentiating between searches for so-called pulsar-term and Earth-term events.  In Sections 7 and 8, we present the results of our pulsar-term and Earth-term searches, respectively.  In Section 9, we place upper bounds on BWM rates and amplitudes. In Section 10 we summarize our key results and offer concluding remarks. We provide a summary of our important notational elements in a table of key symbols in an appendix.

\footnotetext[18]{Department of Physics and Astronomy, Oberlin College, Oberlin, OH 44074, USA}
\footnotetext[19]{Department of Physics and Astronomy, University of New Mexico, Albuquerque, NM 87131, USA}
\footnotetext[20]{Jansky Fellow}
\footnotetext[21]{Einstein Fellow}

%%%%%%%%%%%%%%%%%%%%%%%%%%%%%%%%%%%%%%%%%%%%%%%%%%%%%%%%%%%%%%%%%%%%%%%%%%%
\section{Pulsar Timing Data Set}
In this section, we discuss several aspects of the data that are relevant to our analysis.  For a more thorough description of the data, see \citet{d07} and \citet{dfg+13}.  

The five-year data set consists of pulse times-of-arrival (TOAs) collected at approximately monthly intervals for each of 17 millisecond pulsars (MSPs).  All observations were done with the Arecibo radio telescope and the Green Bank Telescope (GBT).  Of these 17 MSPs, those visible to Arecibo were observed with Arecibo; all others were observed with the GBT.  One pulsar, J1713$+$0747, was observed with both telescopes.  All observations were done using one of two identical backend systems: the Astronomical Signal Processor (ASP) and the Green Bank Astronomical Signal Processor (GASP).  These backends performed real-time coherent dedispersion over bands up to 64 MHz wide and recorded the results averaged over channels of width 4 MHz each. 

For each observing epoch, several TOAs are reported from various frequency channels of the 64 MHz band.  At Arecibo, observations were typically conducted at two widely-separated frequencies (usually 430 MHz and 1400 MHz) within one day; at the GBT, observations at two frequencies (usually 820 MHz and 1400 MHz) were conducted within several days of each other.  Approximately contemporaneous observations at multiple frequencies allow epoch-to-epoch timing fluctuations caused by changes in dispersion measure (DM) to be accounted for in the timing model fit \citep{lcc+14}.  

Three of the pulsars comprising the five-year NANOGrav data set, J1853+1308, J1910+1256, and B1953+29, do not have sufficient dual-frequency coverage to correct for timing errors from variations in DM over time; we exclude these pulsars from our analysis for this reason. We also exclude the data set for J1600$-$3053 from our analysis because it is comparatively very short, spanning just two years.  Searching for BWMs in such short data sets is feasible in principle, but we avoid it for two reasons. First, the minimum detectable BWM amplitude in a particular data set scales approximately as $T^{-3/2}$ where $T$ is the span of the data set \citep{vl10}, so we do not anticipate that this data set will greatly improve our sensitivity to BWMs. Second, in such short data sets, many timing model parameters are highly covariant with each other \citep[e.g., spin and astrometric parameters;][]{mcc13} and with any BWM signal present. 

Finally we exclude the data for J1643$-$1224 from our analysis because we believe it contains chromatic timing biases from unaccounted-for phenomenology in the interstellar medium (ISM; see Figure 2). The DM of J1643$-$1224 is approximately 62 pc cm$^{-3}$, nearly a factor of two greater than any other pulsar in our sample. With high DM pulsars, chromatic timing errors that deviate from the $\nu^{-2}$ scaling expected from cold plasma dispersion alone become more significant \citep{cs10}. Furthermore, this pulsar is directly behind a complex region of H-II associated with $\zeta$-Ophiuchi, a massive, runaway O-type star spinning very near breakup \citep{gmr+01,vh05}. This intervening H-II region could conceivably contribute to non-trivial and currently unaccounted-for chromatic effects on the timing behavior of J1643$-$1224.

%%%%%%%%%%%%%%%%%%%%%%%%%%%%%%%%%%%%%%%%%%%%%%%%%%%%%%%%%%%%%%%%%%%%%%%%%%%
\section{Signal Model}

For a given pulsar, the timing perturbation from a BWM of amplitude $h_B$ is well-modeled as
\begin{eqnarray}
\Delta t(t) = h_BB(\theta,\phi)\left[(t-t_0)\Theta(t-t_0)-\right.\nonumber\\\left.(t-t_1)\Theta(t-t_1)\right].
\end{eqnarray}
The function $B(\theta,\phi)=(1/2)(1-\cos{\theta})\cos{(2\phi)}$ ranges between $-1$ and $1$ and is common to all pulsar timing efforts to detect point-like sources of GWs \citep{ew75,hd83,lwk+11}.  The angle between the direction the burst propagates and the line of sight from Earth to the pulsar is $\theta$; $\phi$ is the angle between the principal polarization vector of the wave and the projection of the line of sight from the Earth to the pulsar onto the plane normal to the wave propagation direction. The BWM encounters the Earth at a time $t_0$ and is observed from Earth to encounter the pulsar at a time $t_1=t_0+(l/c)(1+\cos{\theta})$ where $l$ is the distance from the Earth to the pulsar \citep{vl10,cj12,mcc14}. The function $\Theta$ is the Heaviside step function.  The amplitude of a BWM coming from a SMBHB merger of reduced mass $\mu\equiv M_1M_2/(M_1+M_2)$ ( $M_1$ and $M_2$ are the masses of the black holes in the binary) with a typical inclination angle of ${\cal I}=\pi/3$ at a luminosity distance $D_L$ from Earth is \citep{mcc14}
\begin{eqnarray}
h_B\approx1.5\times10^{-12}\left(\frac{\mu}{10^9~M_\odot}\right)\left(\frac{1~{\rm Mpc}}{D_L}\right).
\end{eqnarray}

The distances to the pulsars in the NANOGrav array are on the order of kiloparsecs.  Our timing baseline is approximately five years.  Unless $\theta$ differs from $\pi$ by less than $\sim$ 3 degrees for a particular pulsar, we expect that $t_0$ and $t_1$ will not both fall within our observing window.  Since we consider only 12 pulsars in our analysis, less than 1\% of the sky is within 3 degrees of one of our pulsars.  Assuming BWMs occur isotropically, there is a less than 1\% chance of both $t_0$ and $t_1$ occurring within our five-year observing span if a BWM occurs at all.  And if a BWM occurred with a small enough angular separation from a pulsar in our array that we could see the timing perturbation both turn on at $t_0$ and turn off at $t_1$, it would only be observed to turn off in that one pulsar. So, in each of our pulsar timing data sets, we need only to look for evidence of timing perturbations of the form
\begin{eqnarray}
\Delta t(t) = h_p(t-t_B)\Theta(t-t_B),
\end{eqnarray}
where $h_p=\pm h_BB(\theta,\phi)$, what we call the projected burst amplitude, and $t_B$ is either $t_0$ or $t_1$, what we call the burst epoch. Bursts arriving at an individual pulsar only influence the timing behavior of that pulsar; we refer to these as pulsar-term bursts (see Section 5). Bursts arriving at the Earth will simultaneously begin to influence the timing behavior of all pulsars in our sample; we refer to these as Earth-term bursts (see Section 6).  
%%%%%%%%%%%%%%%%%%%%%%%%%%%%%%%%%%%%%%%%%%%%%%%%%%%%%%%%%%%%%%%%%%%%%%%%%%%
\section{Noise Model}

Consider TOAs measured using pulse profiles obtained by synchronously averaging ${\cal N}$ pulses for each of several radio-frequency channels.  The pulse profiles are functions of pulse phase, $\varphi$, channel frequency, $\nu$, and observing epoch, $\tau$. A TOA from a particular frequency channel and observing epoch can be written as 
\begin{eqnarray}
t_{\nu,\tau}=t_{\infty,\tau}+t_{{\rm DM}_{\nu,\tau}}+t_{C_{\nu,\tau}}+\epsilon_{{\rm S/N}_{\nu,\tau}}+\epsilon_{J_{\nu,\tau}}+\epsilon_{{\rm DISS}_{\nu,\tau}},\nonumber\\
\end{eqnarray}
where $t_{\infty,\tau}$ is the TOA at infinite frequency, $t_{\rm DM}$ is the dispersive delay from propagation through ionized interstellar plasma, $t_C$ is an additional, non-dispersive chromatic perturbation, such as intrinsic profile evolution with frequency, pulse broadening from multipath propagation, and interstellar refraction \citep{cs10}.  If unaccounted for, $t_{\rm DM}$ and $t_C$ can lead to systematic errors in TOA estimates.  With multi-frequency observations at each observing epoch, effects from $t_{\rm DM}$ are mitigated in the NANOGrav data set.  Potential effects from $t_C$ are combatted by fitting for constant inter-channel offsets unique to each pulsar. Unlike $t_{\rm DM}$ and $t_C$, the `$\epsilon$' terms in Equation 4 are random errors from a variety of noise sources.  

The term $\epsilon_{{\rm S/N}_{\nu,\tau}}$ is uncorrelated between frequency channels and between observing epochs, i.e.,
\begin{eqnarray}
\langle \epsilon_{{\rm S/N}_{\nu,\tau}}\epsilon_{{\rm S/N}_{\nu',\tau'}}\rangle=\sigma^2_{\rm S/N}({\cal N})\delta_{\nu\nu'}\delta_{\tau\tau'},
\end{eqnarray}
where $\sigma_{\rm S/N}\propto(S/N)^{-1}\propto{\cal N}^{-1/2}$.  The quantity $\sigma_{\rm S/N}$ is the TOA uncertainty from radiometer noise assuming that the radiometer noise adds to a fixed pulse shape under the assumptions of matched filtering against a very high S/N template pulse profile.

The second random contribution to $t_{\nu,\tau}$, $\epsilon_{J_{\nu,\tau}}$, is from phase jitter in single pulses.  Jitter is highly correlated between frequency channels, but is known to decorrelate over widely-separated frequencies; the probability distribution function of the phase of single-pulse centroids evolves with frequencies similar to pulse profiles \citep{sod+14}. Jitter noise can be modeled as
\begin{eqnarray}
\langle \epsilon_{J_{\nu,\tau}}\epsilon_{J_{\nu',\tau'}}\rangle=\sigma_J^2({\cal N})\delta_{\tau\tau'}\rho_{J_{\nu,\nu'}}\approx\sigma_J^2({\cal N})\delta_{\tau\tau'},
\end{eqnarray}
where $\sigma_J\propto {\cal N}^{-1/2}$. The quantity $\rho_{J_{\nu,\nu'}}$ is approximately unity unless $\nu$ and $\nu'$ are very widely separated.

Usually, $\sigma_J\lesssim\sigma_{\rm S/N}$. Jitter noise begins to dominate radiometer noise only when the S/N of single pulses begins to exceed unity. For our data, collected with ASP and GASP, jitter noise should be significantly subdominant to radiometer noise for all pulsars. Both radiometer and jitter noise are significantly larger than contributions from $\epsilon_{{\rm DISS}_{\nu,\tau}}$, a random error associated with diffractive interstellar scintillation \citep[DISS;][]{c90,cs10}.  In a detailed study of J1713+0747, \citet{sc12} found that for single pulses, $\sigma_J\approx27\pm1$ $\mu$s, while timing errors from DISS were only a few nanoseconds. A recent 24-hour study of J1713+0747 \citep{dlc+14} confirmed the jitter measurement of \citet{sc12}. Errors from DISS can be substantially more important in the timing of high DM pulsars like B1937+21. We expect DISS to be a subdominant component of our timing error budget for the MSPs we include in our sample, so we ignore it in our analysis.  In summary, the noise covariance matrix for a pulsar in our array, with these anticipated sources of noise, can be approximated as
\begin{eqnarray}
{\bf C}_{\nu\nu',\tau\tau'}=\langle{\epsilon_{\nu,\tau}\epsilon_{\nu',\tau'}}\rangle\approx\delta_{\tau\tau'}\left[\delta_{\nu\nu'}\sigma^2_{\rm S/N}({\cal N})+\sigma_J^2({\cal N})\right],\nonumber\\
\end{eqnarray}
where $\epsilon_{\nu,\tau}$ is the sum of all noise influencing the TOA from observing epoch $\tau$ and frequency channel $\nu$.

\subsection{Empirical Noise Models}

In a previous analysis of the first five years of NANOGrav data, \citet{abb+14} searched for continuous GWs using a different noise model:
\begin{eqnarray}
{\bf C}_{\nu\nu',\tau\tau'}&=&\delta_{\tau\tau'}\left[\delta_{\nu\nu'}({\cal Q}^2+E^2\sigma_{\rm S/N}^2({\cal N}))+{\cal J}^2\right]+\nonumber\\&&{\bf C}_{\rm red}(A,\gamma).
\end{eqnarray}
The terms ${\cal Q}$ and $E$ are commonly referred to as EQUAD (a source of Gaussian white noise with time units added in quadrature to radiometer noise) and EFAC (a dimensionless constant multiplier on the anticipated amount of radiometer noise), respectively.  The term ${\cal J}$ mimics the correlations induced between TOAs from different frequency channels of the same observing epoch by jitter, but qualitatively differs from jitter in that its value does not change to reflect the number of pulses, ${\cal N}$, averaged together to yield a TOA.  Most observations in the five-year NANOGrav data set we considered had similar integration times, so ${\cal N}$ does not fluctuate substantially between epochs and ${\cal J}$ is thus very much like a jitter-induced correlation.  Different values of ${\cal Q}$, $E$, and ${\cal J}$ were used for each widely separated observing frequency of each pulsar.

The \citet{abb+14} noise model also included red noise with a power-law power spectrum $P(f)=A[f/(1~{\rm yr}^{-1})]^\gamma$.  This component of the noise model is not to be thought of as part of the TOA error budget; TOAs measured with extreme accuracy may still differ from the expectations of a timing model because of spin noise intrinsic to a pulsar, which is well modeled as a stochastic process with a power-law spectrum \citep{sc10}.  In the \citet{abb+14} analysis, under the assumption that no GW signal was present in the data set, the noise model parameters ${\bf \Xi}=[A,\gamma,E,{\cal Q},{\cal J}]$ were determined by finding the maximum of the likelihood function
\begin{eqnarray}
{\cal L}({\bf R}|{\bf \delta p},{\bf \Xi})=\frac{\exp{\left[-\frac{1}{2}({\bf R}-{\bf M}{\bf \delta p})^T{\bf C}^{-1}({\bf R}-{\bf M}{\bf \delta p})\right]}}{\sqrt{(2\pi)^{N_{\rm TOA}}\det{{\bf C}}}},\nonumber\\
\end{eqnarray}
where ${\bf R}$ are the timing residuals from an initial timing model, ${\bf \delta p}$ are deviations of the timing model parameters from the initial timing model, ${\bf M}$ is the timing model design matrix, and $N_{\rm TOA}$ is the total number of TOAs in the data set. 

We applied the noise model assessment done by \citet{abb+14} to two simulations of the five-year NANOGrav data set for PSR J1909$-$3744, one that contained only simulated radiometer noise and one that contained radiometer noise and a bright BWM occurring at the midpoint of the data set.  The resulting best-fit noise models were identical except for in the red noise parameters $A$ and $\gamma$. The BWM signal is heavily covariant with the red noise component of the noise model, resulting in a large value for $A$ and a poorly constrained value of $\gamma$. 

An independent analysis of the five-year NANOGrav data set by \citet{pjl+13} found that except for J1643$-$1224 and J1910$+$1256, pulsars we already excluded from our analysis, all pulsars were consistent with white noise alone. Because the demonstrated covariance between red noise and BWM signatures would complicate detection of a BWM \citep{cj12} and because of the lack of strong evidence for red noise, we have opted to consider the \citet{abb+14} fixed noise models without the red noise component.

\begin{figure}
\includegraphics[scale=.46]{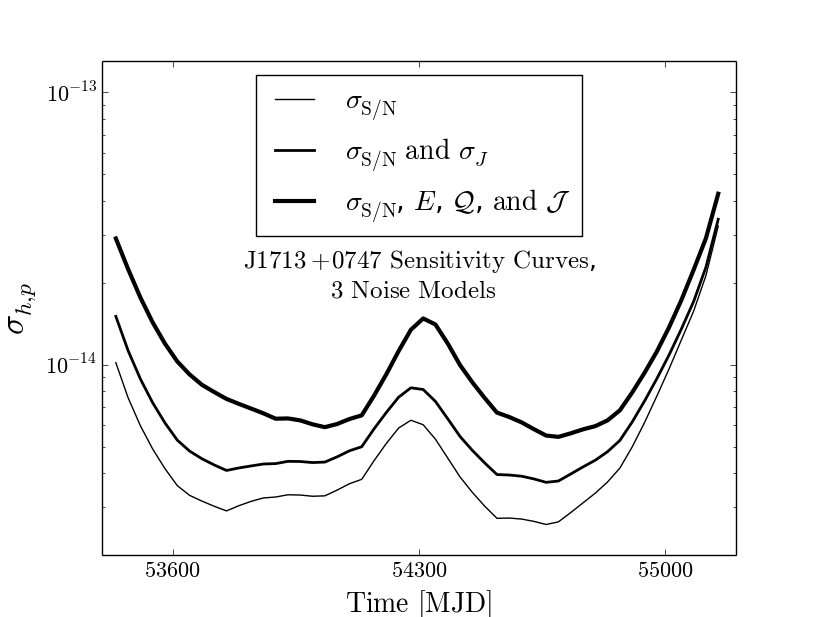}
\caption{Uncertainty on the projected amplitude of a BWM at various trial burst epochs with the NANOGrav five-year data set for PSR J1713$+$0747 using three different noise models. The lowest curve assumes that only radiometer noise is present in the data and is identical to the results of \citet{mcc14}. The middle curve assumes that only radiometer noise and jitter noise are present in the data (as in Equation~7); the scale of the jitter contribution is consistent with \citet{sc12}.  The most conservative curve is based on the fixed noise model used in \citet{abb+14} (as in Equation~8) sans a red noise component.}
\end{figure}

\subsection{Comparing Noise Models}
In Figure 1, we show how the uncertainty on the projected BWM amplitude, $\sigma_{h,p}$, varies in the NANOGrav five-year data set for PSR J1713+0747 over several trial burst epochs under three different noise models. The quantity $\sigma_{h,p}$ is a direct measure of how sensitive a particular data set is to BWMs and it is directly influenced by the noise model; it is discussed by, e.g., \citet{vl10} and \citet{mcc14}.

The bottom curve of Figure 1 is based on a noise model that only includes radiometer noise. The middle curve uses the noise model described by Equation~7 with the jitter measurement from \citet{sc12}.  The top curve is based on the noise model used by \citet{abb+14} and described in Equation~8 (without the red noise component). The discrepancy between the physically motivated noise model of Equation~7 and the empirical noise model of Equation~8 is not currently well understood. Detailed studies like \citet{dlc+14} are being carried out to better understand the noise budget of pulsar timing experiments and any systematic effects that influence the timing procedure, but bridging the gap between the top and middle curves of Figure 1 is an ongoing area of research. The \citet{abb+14} noise model is the most conservative of the three we consider so we use it for the remainder of our analysis.

%%%%%%%%%%%%%%%%%%%%%%%%%%%%%%%%%%%%%%%%%%%%%%%%%%%%%%%%%%%%%%%%%%%%%%%%%%%
\subsection{Epoch Averaging}

\begin{figure}
\includegraphics[height=150mm,width=95mm]{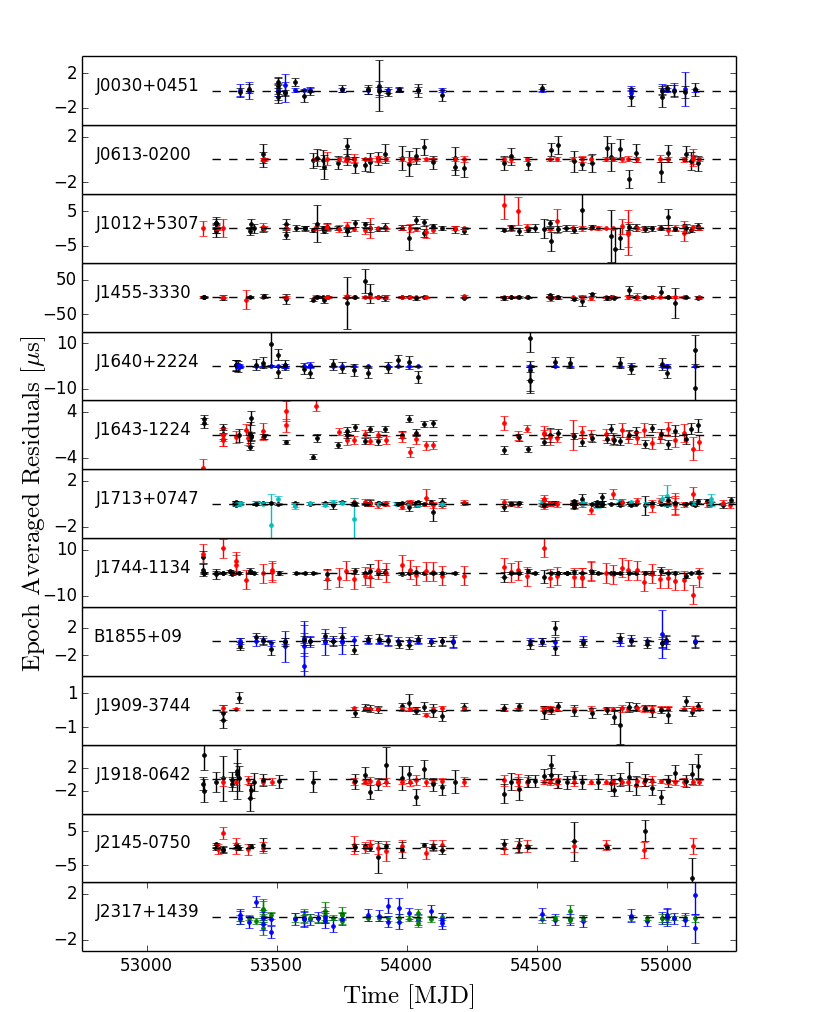}
\caption{Epoch-averaged timing residuals for the 12 pulsars used in our analysis (and the excluded pulsar J1643$-$1224). The various colors indicate different observing frequencies: 327 MHz is green, 430 MHz is blue, 820 MHz is red, 1400 MHz is black, and 2300 MHz is cyan. We show the residuals for J1643$-$1224 to illustrate the apparent chromatic problems with the data: the residuals from 820 MHz and 1400 MHz are consistently anti-correlated and are often extreme outliers.}
\end{figure}

In Figure 2, we depict the epoch-averaged timing residuals from the 12 pulsars in our sample (and the excluded pulsar J1643$-$1224). Residuals are obtained individually for multiple frequency channels for each integration lasting 15 to 45 minutes. The residuals from that integration are combined to form a single epoch-averaged residual. The noise model is central to this procedure. We define an operator,
\begin{eqnarray}
{\bf A} = \left({\bf U}^T{\bf C}^{-1}{\bf U}\right)^{-1}{\bf U}^T{\bf C}^{-1},
\end{eqnarray}
which maps the raw residuals, ${\bf R}$, to epoch-averaged residuals, ${\bf R}_E={\bf AR}$. The matrix ${\bf U}$ is the ``exploder" matrix discussed in \citet{abb+14} that maps epochs to the full set of TOAs. The uncertainties on the epoch-averaged residuals are the square roots of the diagonal entries of the matrix
\begin{eqnarray}
{\bf C}_E=\langle {\bf R}_E{\bf R}_E^T\rangle= \left({\bf U}^T{\bf C}^{-1}{\bf U}\right)^{-1}.
\end{eqnarray}

%%%%%%%%%%%%%%%%%%%%%%%%%%%%%%%%%%%%%%%%%%%%%%%%%%%%%%%%%%%%%%%%%%%%%%%%%%%
\section{Searches for BWMs in the Pulsar Term}

Information in the pulsar terms of the 12 pulsars in our sample comes from causally distinct regions of spacetime \citep{cj12}.  A BWM is 12 times more likely to encounter a single pulsar in our array than it is to encounter the Earth.  However, it is not true that BWMs encountering individual pulsars in the array are 12 times more likely to be confidently detected.  Different pulsars are timed with varying degrees of precision and with differing observing cadences making them unequally sensitive in searches for BWMs.  A particular BWM may be polarized in such a way or from such a part of the sky that the projection factor, $B(\theta,\phi)$, makes its influence on the timing behavior of a particular pulsar vanishingly small.  Furthermore, non-BWM phenomena such as intrinsic pulsar spin noise \citep{sc10} and microglitches \citep{cb04} can be confused as BWMs. Without the signal appearing concurrently in multiple pulsars as in an Earth-term BWM, it is difficult to rule out pulsar-specific phenomena. Nonetheless, we carry out a search for BWMs in each of our individual pulsars as non-detections in many pulsars allow us to place constraints on otherwise inaccessible regions of BWM parameter space.

\citet{mcc14} describe techniques for searching for BWMs in individual pulsar timing data sets and for assessing the minimal projected amplitude detectable at a particular epoch.  The timing perturbation from a BWM is deterministic and can be included as part of the timing model.  The projected amplitude enters the timing model as a linear parameter and can be fit for using least-squares methods \citep{g10}, yielding an estimate for the projected amplitude, $\hat{h}_p$, and its uncertainty, $\sigma_{h,p}$.  We deal with the nonlinear parameter $t_B$ by searching over a grid of trial burst epochs. For all of our timing model fits and calculation of timing residuals, we use the software package \texttt{TEMPO2} \citep{ehm06}.

The modification to the timing solution caused by including a burst of projected amplitude $h_p$ at time $t_B$ can be assessed by computing the likelihood ratio for a model with and without a burst:
\begin{eqnarray}
\Gamma(h_p,t_B)=\exp{\left(-\frac{1}{2}\left[\chi^2(h_p,t_B)-\chi^2(0,t_B)\right]\right)}.
\end{eqnarray}
In this expression,
\begin{eqnarray}
\chi^2(h_p,t_B)={\bf R}^T(h_p,t_B){\bf C}^{-1}{\bf R}(h_p,t_B), 
\end{eqnarray}
where ${\bf R}(h_p,t_B)$ are the timing residuals when a burst of projected amplitude $h_p$ at time $t_B$ is included as part of the timing model. The least-squares estimator for the projected burst amplitude at a trial burst epoch $t_{B}$, $\hat{h}_p(t_{B})$, maximizes $\Gamma$ at that trial burst epoch; call this value $\hat{\Gamma}(t_{B})$. For a fixed trial burst epoch, if the data are consistent with the noise model, $\hat{D}\equiv2\ln{\hat{\Gamma}}$, the reduction in the $\chi^2$ of the residuals caused by introducing a projected burst amplitude to the timing model fit will be a random variable following a $\chi^2$ distribution with one degree of freedom:
\begin{eqnarray}
f_1(\hat{D})=(2\pi\hat{D})^{-1/2}\exp{(-\hat{D}/2)}.
\end{eqnarray}
Since the burst amplitude is a linear parameter in the timing model, we expect the $\chi^2$ value of the residuals to respond quadratically to the burst amplitude, i.e.,
\begin{eqnarray}
\chi^2(0,t_B)=\chi^2\left(\hat{h}_p(t_B),t_B\right)+\left[\hat{h}_p(t_B)/\sigma_{h,p}(t_B)\right]^2,\nonumber\\
\end{eqnarray}
or $\hat{D}(t_B)=[\hat{h}_p(t_B)/\sigma_{h,p}(t_B)]^2$. The $\sigma_{h,p}$ values are the least-squares 1-$\sigma$ amplitude uncertainties on $\hat{h}_p$, e.g., the quantities plotted in Figure 1. 

If we compute $\hat{D}$ for two trial epochs that are very close together, the results will be correlated. Because of these correlations, when computing $\hat{D}$ along a densely sampled grid of $N_t$ trial burst epochs, the number of effectively independent trial burst epochs tested is $N_I<N_t$. The probability distribution function for the maximum value of $\hat{D}$ along the grid, $\hat{D}_{\rm max}$, is 
\begin{eqnarray}
f_{N_I}(\hat{D}_{\rm max})=N_If_1(\hat{D}_{\rm max})F_1(\hat{D}_{\rm max})^{(N_I-1)},
\end{eqnarray}
where $F_1$ is the cumulative distribution of $f_1$,
\begin{eqnarray}
F_1(\hat{D})={\rm erf}{\left(\sqrt{\hat{D}/2}\right)}.
\end{eqnarray}
The cumulative distribution for $\hat{D}_{\rm max}$ associated with $f_{N_I}$ is then simply
\begin{eqnarray}
F_{N_I}(\hat{D}_{\rm max})={\rm erf}^{N_I}\left(\sqrt{\hat{D}_{\rm max}/2}\right).
\end{eqnarray}
The anticipated false alarm probability for noise alone to exceed a threshold value of $\hat{D}_{\rm max}$, $\hat{D}_{\rm thresh}$, is just
\begin{eqnarray}
{\cal F}_{N_I}(\hat{D}_{\rm thresh})=1-F_{N_I}(\hat{D}_{\rm thresh}). 
\end{eqnarray}
For a fixed allowable false alarm probability, $\hat{D}_{\rm thresh}$ grows logarithmically with $N_I$ if $N_I \gg 1$.

\begin{figure}
\includegraphics[height=70mm]{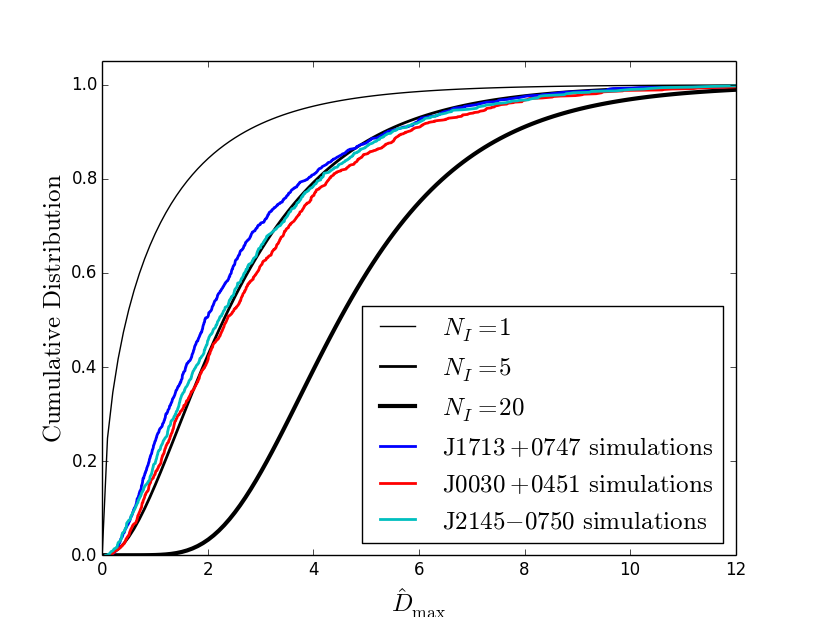}
\caption{Cumulative distributions of $\hat{D}_{\rm max}$ values from 1000 simulations of noise-like timing residuals for three of the pulsars in our sample. Overlaid are the anticipated cumulative distributions if $N_I$, the effective number of trial burst epochs tested, is one, five, or twenty. For all pulsars in our sample, the analytically anticipated cumulative distribution best fits the results of our simulations with $N_I$ between 4 and 5.}
\end{figure}

To estimate $N_I$, for each pulsar, we generated 1000 simulated sets of TOAs that matched the real data set in number of TOAs, observing schedule, and timing model, but yielded timing residuals consistent with our noise models. We then fit for $\hat{h}_p(t_{B})$ along an equispaced grid of twenty trial burst epochs between MJD 53500 and 54900 and computed $\hat{D}(t_{B})$ to get $\hat{D}_{\rm max}$. 

In Figure 3, we show the cumulative distribution of the 1000 $\hat{D}_{\rm max}$ values from our simulations for three pulsars: J0030$+$0451 and J2145$-$0750, observed by Arecibo and the GBT, respectively, with rms timing residuals of $\sim$~100~ns, and J1713$+$0747, observed by both Arecibo and the GBT with an rms timing residual of $\sim$~30~ns.  We also plot the theoretically anticipated curves for $N_I$ equal to one, five, and twenty. In fitting the anticipated cumulative distributions of $\hat{D}_{\rm max}$ with $N_I$ as a free parameter to the results of our simulations, we find that for all pulsars, $N_I$ is between four and five. As $N_I$ increases, large values of $\hat{D}_{\rm max}$ occur more frequently in the presence of pure noise. We take the conservative approach and assume that there are five independent trial burst epochs to test. With $N_I$ fixed at five, inverting Equation~19 allows us to compute $\hat{D}_{\rm thresh}$ for any desired allowable false alarm probability; false alarm probabilities of approximately five and one percent are expected for $\hat{D}_{\rm thresh}$ equal to 6.60 and 9.54, respectively.

%%%%%%%%%%%%%%%%%%%%%%%%%%%%%%%%%%%%%%%%%%%%%%%%%%%%%%%%%%%%%%%%%%%%%%%%%%%
\section{Searches for BWMs in the Earth Term}
Searches for BWMs in the Earth term have several advantages over searches in pulsar terms. The timing perturbation from a BWM will turn on simultaneously for all pulsars in the PTA if it occurs in the Earth term, providing a powerful means by which pulsar-specific phenomena (e.g., glitches) can be ruled out. Variations in the projected BWM amplitude from pulsar to pulsar provide information about the location of the GW source and its polarization.  As such, for a trial BWM source location, polarization, and epoch, the residuals of all pulsars in the PTA can be combined into a coherent global fit for the true burst amplitude, $h_B$. Such global fits, described in \citet{mcc14}, have better amplitude sensitivity than what is attainable with any one timing data set and can be carried out with a similar least-squares apparatus as is used in pulsar term searches.  

Global least-squares fitting techniques were recently used by the PPTA to search for BWMs \citep{whc+14}. The PPTA data set those authors analyzed contained many fewer TOAs than the NANOGrav data set we are considering here owing to NANOGrav's practice of reporting many TOAs from a single observing epoch but from different frequency channels.  With these multifrequency TOAs, NANOGrav includes as part of its timing models numerous chromatic parameters to account for profile evolution across our wide-bandwidth receivers and time-variable DM, so, we fit for many more timing model parameters than the PPTA. 

Suppose $N$ is the number of timing model parameters being fit for each of $M$ pulsars in a PTA (in reality, $N$ varies from pulsar to pulsar). Then carrying out a global fit for the amplitude of a BWM requires the inversion of an $(NM+1)\times (NM+1)$ matrix, a procedure that requires computation time $\propto (NM+1)^3$.  This global fit has to be done over $N_{\Omega}N_{\psi}N_t$ trials in the four-dimensional phase space of sky position, burst epoch, and burst polarization ($N_\Omega$, $N_\psi$, and $N_t$ are, respectively, the number of sky positions, polarization angles, and burst epochs tested).  Methods requiring these global fits are thus computationally onerous for NANOGrav because of the large number of timing model parameters for which we fit. More importantly, the total number of suitably stable MSPs being timed by NANOGrav is growing with time as projects like the Arecibo PALFA survey \citep[e.g.,][]{cfl+06,slm+14}, the Arecibo All-sky 327-MHz Drift Pulsar Survey \citep[e.g.,][]{dsm+13}, and the GBT Northern Celestial Cap Pulsar Survey \citep[e.g.,][]{slr+14} continue to find more pulsars. A search procedure based on global fits will not scale well to future PTA endeavors. We use a different technique that avoids having to carry out any global fits and is computationally much more efficient.

\subsection{Accelerated Earth-term Searches}
Over a five-dimensional grid of trial burst sky positions, $\Omega_{ij}$, polarizations, $\psi_k$, epochs, $t_{B,l}$, and amplitudes, $h_{B,m}$, we search for the maximum in the global likelihood ratio surface,
\begin{eqnarray}
\Gamma_G(\Omega_{ij},\psi_k,t_{B,l},h_{B,m})&&\\\nonumber=&&\prod_{K=1}^M\Gamma_K[B_K(\Omega_{ij},\psi_k)h_{B,m},t_{B,l}],
\end{eqnarray}
where $B_K$ is the projection factor, $B(\theta,\phi)$, for the $K$th pulsar. The $\Gamma_K$ quantities are analogous to the likelihood ratios in Equation 12. The angles $\theta$ and $\phi$ depend on the location of the $K$th pulsar and the trial burst location, $\Omega_{ij}$. The trial burst polarization angle, $\psi_k$, influences $\phi$. The total number of trials in this five-dimensional search is $N_\Omega N_\psi N_tN_h$, where $N_h$ is the number of trial burst amplitudes tested.

Ostensibly, the procedure appears to require that for each grid point in the five-dimensional space and for each pulsar in our sample, we incorporate the signature of a BWM of projected amplitude $B_K(\Omega_{ij},\psi_k)h_{B,m}$ occurring at time $t_{B,l}$ into the timing model of the $K$th pulsar, refit the timing model, and compute $\Gamma_K$. This would take computation time $\propto N_\Omega N_\psi N_tN_hMN^3$, a speed up over the global fitting scheme so long as $N_h<(NM+1)^3/MN^3$. The average $N$ for our sample of $M=12$ pulsars is approximately 90, meaning $N_h$ needs to be less than approximately 140 for this technique to be faster. 

However, there is a greater speed-up to be had. Variations in $\Omega_{ij}$, $\psi_k$, and $h_{B,m}$ only alter the projected BWM amplitude along different pulsar lines of sight. We can precompute the two-dimensional (projected burst amplitude and burst epoch) likelihood ratio surface for each pulsar in our sample, a procedure requiring computation time $\propto N_pN_tMN^3$, where $N_p$ is the number of trial projected burst amplitudes considered. We then construct the full five-dimensional likelihood ratio surface from Equation~20 by figuring out what the projected burst amplitude in the $K$th pulsar would be for a particular choice of $\Omega_{ij}$, $\psi_k$, and $h_{B,m}$, looking to the precomputed two-dimensional likelihood ratio surface for the $K$th pulsar, and interpolating between the nearest projected burst amplitudes tested and the projected amplitude of interest from the five-dimensional search; this step requires no additional timing model fits (the computationally costly step), is very fast, and is essentially an exercise in efficiently searching lookup tables.  

Using precomputed two-dimensional likelihood ratio surfaces for a global Earth-term BWM search as we have just described will lead to a speed up over the global fitting scheme if $N_p<N_\Omega N_\psi (NM+1)^3/MN^3$. For our search, we have tested $N_t=40$ trial burst epochs evenly spaced between MJDs 53541 and 54995, $N_\psi=17$ trial burst polarization angles evenly spaced between $0$ and $\pi$, and $N_\Omega=1598$ trial burst sky positions isotropically distributed on the sky. As a brief but important aside, we have chosen to search within the particular window of dates just mentioned because for each pulsar in our sample, there is at least one collection of multifrequency observations before and after this window, with the limiting pulsar on the early side being J0613$-$0200 and J2145$-$0750 on the late side. With the grid parameters we have chosen, so long as $N_p$ is less than approximately $4\times10^6$, using precomputed two-dimensional likelihood ratio surfaces will be faster than global fitting schemes. In practice, we have set $N_p=300$, testing 150 trial projected amplitudes logarithmically spaced between $5\times 10^{-17}$ and $10^{-12}$ and their negatives. Utilizing the precomputed two-dimensional likelihood ratio surfaces is thus faster than using global fitting by a factor of over thirteen thousand and is the only reason why searching such a densely sampled grid is feasible.

Once the global, five-dimensional likelihood ratio surface is computed, for fixed $\Omega_{ij}$, $\psi_k$, and $t_{B,l}$, we isolate the value of $h_{B,m}$ that maximizes $\Gamma_G$; we call it $\tilde{h}_B$. This four-dimensional surface, which we call $\tilde{\Gamma}_G$, is approximately equal to $\hat{\Gamma}_G$, what we would get if we had carried out the computationally costly global fits we have taken great care to avoid, differing from $\hat{\Gamma}_G$ only because of the finite resolution of our trial burst amplitude grid. We can then, as in the pulsar-term case, consider the false alarm statistics of the quantity $\tilde{D}_G=2\ln{\tilde{\Gamma}_G}$. For any fixed grid point in our four-dimensional parameter space, if the data are consistent with our noise models, $\tilde{D}_G$ again follows $\chi^2$ statistics with one degree of freedom. With noise-like data, if we consider the probability distribution of $\tilde{D}_{G,{\rm max}}$, the maximum value of $\tilde{D}_G$ over the whole grid, also follows Equation~16, but with a larger number of effective independent trials sampled than in the pulsar-term case owing to a search not just over time, but also over polarization angle and sky position. We call the effective number of independent trials in the global search $N_G$. We discuss $N_G$ in more detail in Section~8.

\subsection{Assessing BWM Amplitude Uncertainty}
A similar relationship exists between $\tilde{D}_G$, $\tilde{h}_B$, and $\sigma_h$ as exists between $\hat{D}$, $\hat{h}_p$ and $\sigma_{h,p}$ as discussed in Section 5, i.e., $\hat{D}=(\hat{h}_p/\sigma_{h,p})^2$. Since $\Gamma_G$ is just the product of all the pulsar-specific likelihood ratio surfaces (with appropriate projection factors, as in Equation~20),
\begin{eqnarray}
\tilde{D}_G = \sum_{K=1}^MD_K(B_K\tilde{h}_B)=\tilde{h}_B^2\sum_{K=1}^M(B_K/\sigma_{h,p,K})^2.
\end{eqnarray}  
So, we have 
\begin{eqnarray}
\tilde{D}_G = (\tilde{h}_B/\sigma_h)^2,
\end{eqnarray}
where
\begin{eqnarray}
\sigma_h = \left[\sum_{K=1}^M(B_K/\sigma_{h,p,K})^2\right]^{-1/2}.
\end{eqnarray}
Equation 23 is equivalent to Equation 32 from \citet{vl10}. While those authors analyzed idealized pulsar timing data sets with equispaced observing epochs, uniform TOA uncertainties, and very simple timing models, their result holds more generally and applies here.

%%%%%%%%%%%%%%%%%%%%%%%%%%%%%%%%%%%%%%%%%%%%%%%%%%%%%%%%%%%%%%%%%%%%%%%%%%%
\subsection{Cross-checks with Bayesian Methods}

As an independent check, we also carry out the Bayesian method of \citet{vl10}, implemented in the software package {\rm
Piccard}\footnote[21]{https://github.com/vhaasteren/piccard}, developed
independently of the \citet{mcc14} method. In brief, this Bayesian
method takes the likelihood of Equation 9 as a function of the noise parameters
${\bf \Xi}$, the timing model parameters ${\bf \delta p}$, and the BWM
parameters $(\Omega_{ij},\psi_{k},t_{B,l}, h_{B,m})$. Similar to what we do
for our frequentist search, we keep the noise parameters fixed
to the values obtained in \citet{abb+14}. Our prior distributions
are flat in all stated BWM parameters, where we note that the prior in
$\Omega_{ij}$ is taken flat over the sphere. We have been able to confirm that
the results we describe in this paper are identical for the frequentist and
Bayesian methods.

%%%%%%%%%%%%%%%%%%%%%%%%%%%%%%%%%%%%%%%%%%%%%%%%%%%%%%%%%%%%%%%%%%%%%%%%%%%
\section{Pulsar Term Results}

For all twelve of the pulsar data sets we analyze, our search for pulsar-term BWMs occurring between MJDs 53541 and 54994 yields results that are entirely consistent with our noise models. In Table 1, we summarize the key results in our search for pulsar-term BWMs. For each pulsar in our sample, we list the pulsar name, the most significant value of $\hat{h}_p$ (according to the $\hat{D}$ number) along with its 1-$\sigma$ uncertainty, the trial burst epoch at which this most significant amplitude was found, the $\hat{D}_{\rm max}$ value for that pulsar, and the logarithm of the false alarm probability anticipated from noise alone for that value of $\hat{D}_{\rm max}$ assuming $N_I=5$. The most significant event we find is in the data for J1918$-$0642; it is consistent with approximately seven percent of noise realizations.

\begin{deluxetable}{lcccc}
\tablewidth{0pc}
\tablecolumns{5}
\tablecaption{Summary of Results from Pulsar Term BWM Search}
\tablehead{
	\colhead{PSR}&
	\colhead{$\hat{h}_p^{\rm max}/10^{-13}$}&
	%\colhead{}&
	\colhead{$t_{B,i}^{\rm max}$}&
	\colhead{$\hat{D}_{\rm max}$}&
	\colhead{$\log_{10}{({\cal F}_5)}$}\\
	\colhead{}&
	%\colhead{}&
	\colhead{}&
	\colhead{(MJD)}&
	\colhead{}&
	\colhead{}}
\startdata
J0030$+$0451& $-$0.87$\pm$0.77 &54100 &1.2&$-$0.11\\
J0613$-$0200 & $~~~$0.68$\pm$0.50& 54584 &1.8&$-$0.20\\
J1012$+$5307 &$~~~$1.47$\pm$0.79 &54062 &3.4&$-$0.54\\
J1455$-$3330 &$-$2.57$\pm$3.35 &54211 &0.5&$-$0.02\\
J1640$+$2224& $-$2.65$\pm$1.45 &53801 &3.3&$-$0.52\\
J1713$+$0747 &$~~~$0.08$\pm$0.07 &54808 &1.4&$-$0.13\\
J1744$-$1134 &$-$2.06$\pm$1.19 &53541 &2.9&$-$0.45\\
B1855$+$09 &$~~~$1.00$\pm$0.73& 53578 &1.8&$-$0.21\\
J1909$-$3744& $-$1.29$\pm$0.74 &54994 &3.0&$-$0.45\\
J1918$-$0642& $-$8.29$\pm$3.39 &54994 &5.9&$-$1.15\\
J2145$-$0750& $~~$26.60$\pm$11.51 &54994 &5.3&$-$0.99\\
J2317$+$1439 &$-$2.40$\pm$1.20 &53541 &3.9&$-$0.67
\enddata
\tablecomments{Description of most significant BWM-like signal detected in the pulsar term of each of 12 NANOGrav data sets. From left to right, the columns are the name of the pulsar, the most significant projected BWM amplitude calculated from least-squares fitting, the trial burst epoch at which the most significant amplitude was found, the corresponding $\hat{D}_{\rm max}$ value, and the logarithm of the anticipated false alarm probability for that value of ${\hat D}_{\rm max}$ assuming $N_I$, the number of effectively independent trial burst epochs tested, is five.}
\end{deluxetable}

\begin{comment}
J1643$-$1224 &5.475   $\pm$0.715 &54137.3 &58.63&$-$13.022\\
\end{comment}

The epochs at which the most significant projected BWM amplitudes occur are distributed nearly uniformly throughout the range of trial epochs we tested with repeat values occurring only at the very first and last epochs tested. Clustering at the edges of the window of tested epochs is not surprising. If there are few timing residuals outside of the window of tested trial burst epochs, when testing the first or last trial epoch, the pre- or post-burst timing model is constrained by a small number of data points. If the residuals at the edge of the data set are slight outliers, allowing for an instantaneous change in the spin period of the pulsar near the beginning or end of the data set can bring them more in line with zero and lead to a modest reduction of the $\chi^2$ value of the residuals. 

The individual data set in our sample that most tightly constrains BWMs is for J1713$+$0747. This is in line with the expectations of \citet{mcc14}. In Figure 4, we show the detailed BWM search results for J1713$+$0747. The bottom panel is simply the epoch-averaged residuals as in Figure 2; we plot them again here to explicitly show how the interval of trial burst epochs tested overlaps with the TOA coverage. The second panel from the bottom shows $\hat{h}_p\pm\sigma_{h,p}$ at the 40 trial burst epochs we tested. The best-fit projected amplitude is completely consistent with zero at each trial burst epoch considered. Further echoing this, in the third panel from the bottom, we show the value of $\hat{D}$ as it varies with trial burst epoch. Never does $\hat{D}$ approach the values necessary to be inconsistent with noise at the five- or one-percent level. In the remaining panels of Figure 4, we show a ``heat map" of the two-dimensional likelihood ratio surface, as in Equation~12, that we use in our Earth-term search. 

\begin{figure}
\includegraphics[height=113mm]{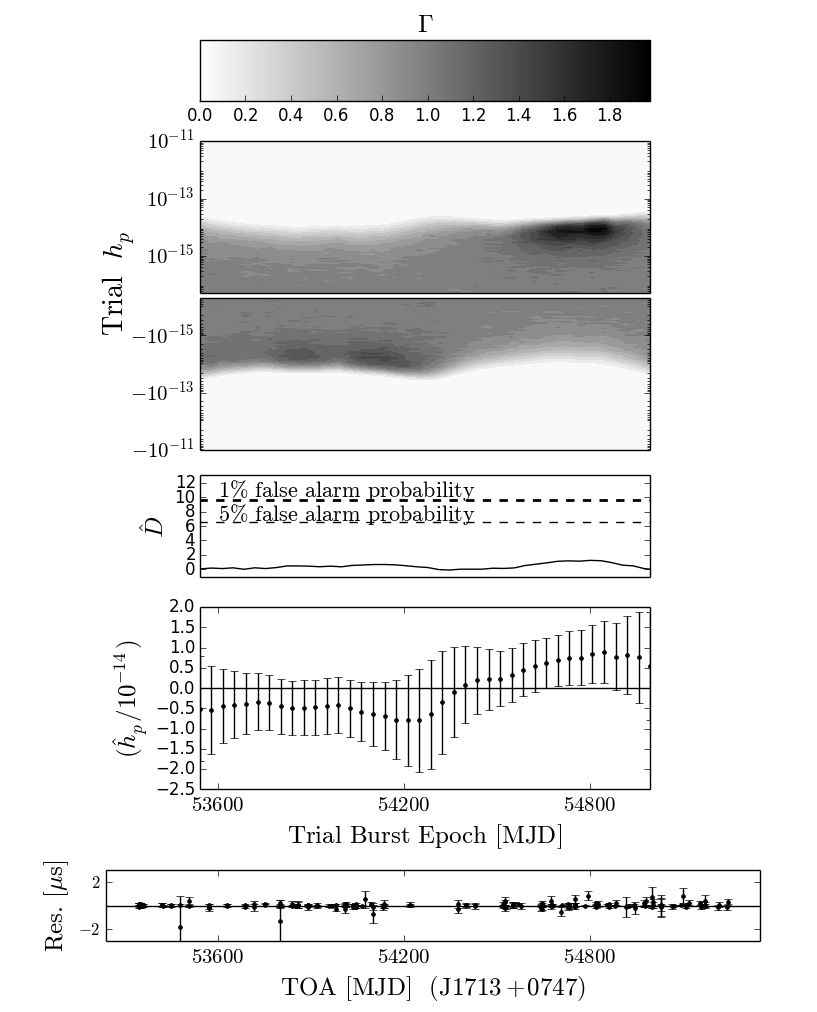}
\caption{Results of a search for a BWM in the NANOGrav five-year data set for PSR J1713$+$0747, the single most sensitive data set in our sample for such searches. The bottom panel shows the epoch-averaged timing residuals for J1713$+$0747 as they are in Figure 2. The second panel from the bottom shows $\hat{h}_p\pm\sigma_{h,p}$ for the 40 trial burst epochs we tested. The third panel from the bottom shows the detection statistic $\hat{D}$ varying over the span of trial burst epochs tested. The remaining panels show a ``heat map" of the two-dimensional likelihood ratio surface, as in Equation 12, that we use for our Earth-term analysis.}
\end{figure}

%%%%%%%%%%%%%%%%%%%%%%%%%%%%%%%%%%%%%%%%%%%%%%%%%%%%%%%%%%%%%%%%%%%%%%%%%%%
\section{Earth Term Results}

\begin{figure}
\begin{center}
\includegraphics[height=55mm,width=100mm]{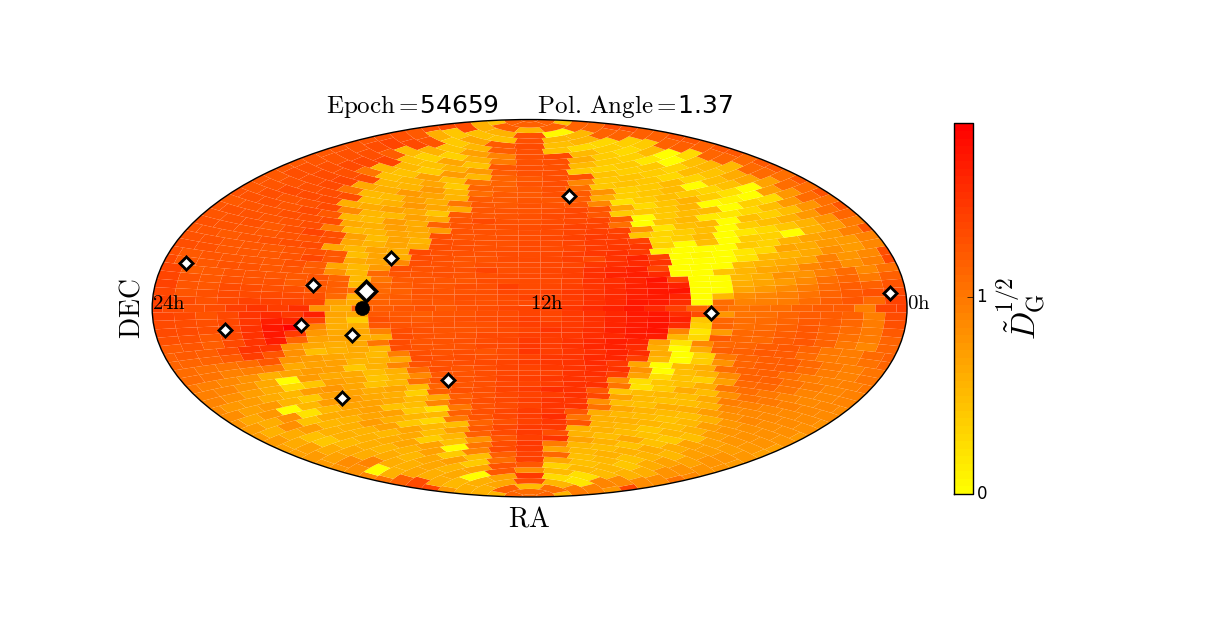}
\end{center}
\caption{A time and polarization slice of our Earth-term search. The plotted quantity, $\tilde{D}_G^{1/2}$, is equivalent to the best-fit BWM amplitude, $\tilde{h}_B$, in units of the 1-$\sigma$ uncertainty on the amplitude, $\sigma_h$. At the location indicated by the black circle, $\sigma_h=6.07\times 10^{-15}$, the smallest value of $\sigma_h$ in our entire search. In our whole search, the maximum value of $\tilde{D}_G^{1/2}$ we find is approximately 2.46, entirely consistent with our noise models at the 95\% level even if $N_G$ is no greater than 5 as we used in our pulsar-term searches. The diamonds indicate the positions of the 12 pulsars in our analysis. The largest diamond represents J1713$+$0747.}
\end{figure}

\begin{figure}
\begin{center}
\includegraphics[height=55mm,width=100mm]{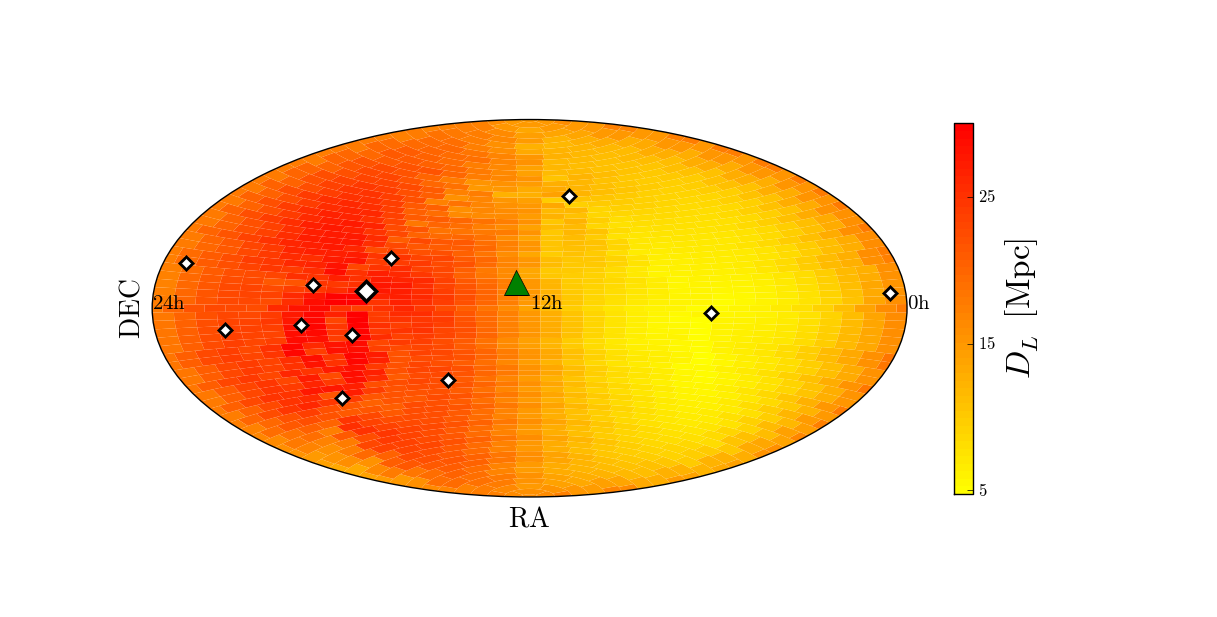}
\end{center}
\caption{The maximum luminosity distance of a SMBHB merger causing a BWM detectable with 95\% confidence given our sensitivity averaged over burst epoch and polarization angle. We have assumed the binary had a typical inclination angle of $\pi/3$ and a reduced mass of $10^9~M_\odot$; the plotted distances scale linearly with this fiducial reduced mass. The diamonds indicate the positions of the 12 pulsars in our analysis. The largest diamond represents J1713$+$0747. The green triangle represents the position of the center of the Virgo Cluster. Just 16.5 Mpc from Earth, the Virgo Cluster is near the edge of the volume in which we can detect BWMs with these properties.}
\end{figure}

In Figure 5, we show a two-dimensional slice of our Earth-term search results. For fixed trial burst polarization angle and epoch, we depict the quantity $\tilde{D}_G^{1/2}$ as it varies with trial burst source location. The quantity $\tilde{D}_G^{1/2}$ is the best-fit BWM amplitude, $\tilde{h}_B$, in units of the uncertainty on the amplitude, $\sigma_h$ (see Equation 22). The slice we show contains the smallest value of $\sigma_h$ found in our entire search, i.e., the point in our parameter space where we are most sensitive to a BWM; its location on the sky is indicated by the black circle very near the location of J1713$+$0747.  

The single largest value of $\tilde{D}_G$ we find is $\tilde{D}_{G,{\rm max}}=6.03$, which corresponds to $\tilde{h}_G = 2.46\sigma_h$. We mentioned at the end of Section 6.1 that the effective number of trials tested in an Earth-term BWM search, $N_G$, will be larger than $N_I=5$ because of the search over not just many trial burst epochs, but also over many trial burst source locations and polarizations, implying that for a fixed allowable false-alarm probability we have to raise the threshold we impose on $\tilde{D}_G$ above the threshold used on the detection statistic in a pulsar-term search. However,   $\tilde{D}_{G,{\rm max}}$ is small enough that even if $N_G$ were only five, this result would be entirely consistent with more than 95\% of realizations of our noise. 

In the BWM analysis carried out by \citet{whc+14}, when faced with marginally high values of their test statistic, they conducted an extensive suite of simulations to assess $N_G$, comparable to what we have done to assess $N_I$ (and depicted in Figure 3), and were able to justifiably raise the detection threshold on their test statistic and rule out a detection. We find no comparably large values of $\tilde{D}_G$ that exceed our detection threshold even if we underestimate $N_G$ as five.

We do still want an estimate of $N_G$ as it will allow us to apply upper limits to the population of BWMs given our non-detection. \cite{cv14} recently demonstrated that the response of a PTA to GWs of any waveform can be decomposed into a linear combination of a finite number of modes, or sky maps. The number of modes required is equal to twice the number of pulsars in the array (the factor of two accounts for the two possible polarization modes of GWs). We use their result to estimate that the number of statistically independent samples in our Earth-term search over source-position and polarization space is 24, twice the number of pulsars used in our analysis. Given five independent samples in time, we thus adopt $N_G=120$. 

Adopting $N_G=120$ is a conservative estimate as our sensitivity to BWMs is so strongly dominated by a single pulsar, so the effective number of pulsars in our array is fewer than 12. As mentioned following Equation~19, for a fixed allowable false alarm probability, if $N_G\gg 1$, $\tilde{D}_{G,{\rm thresh}}$ only diverges logarithmically with $N_G$, so overestimates of $N_G$ are not exceedingly deleterious for the purpose of setting upper limits. With $N_G=120$, setting $\tilde{D}_{G,{\rm thresh}}=12.4$ assures a false alarm probability of less than five percent. This is equivalent to requiring that $\tilde{h}_B$ be greater than $3.52\sigma_h$ in order for it to be inconsistent with 95\% of realizations of noise. Again, in our Earth-term search, we find no signal that meets or exceeds this level of significance.

In Figure 6, we have averaged $\sigma_h$ over trial burst polarization angles and epochs and shown the maximum luminosity distance at which a SMBHB merger with an inclination angle of $\pi/3$ and $\mu=10^9~M_\odot$ (consistent with Equation 2) would be detectable with our data set with 95\% confidence, or where $\tilde{h}_B=3.52\langle\sigma_h\rangle_{\psi,t}$. Our sensitivity is worst near the position of PSR J0613$-$0200. This has little to do with J0613$-$0200, but is instead because these sky positions are antipodal to our greatest concentration of pulsars, especially our most sensitive pulsar, J1713$+$0747. The green triangle in Figure 6 indicates the position of the Virgo Cluster. Just 16.5 Mpc from Earth, the Virgo Cluster falls very near the edge of the volume over which we are sensitive to BWMs from $\mu=10^9$ $M_\odot$, ${\cal I}=\pi/3$ binary mergers.

%%%%%%%%%%%%%%%%%%%%%%%%%%%%%%%%%%%%%%%%%%%%%%%%%%%%%%%%%%%%%%%%%%%%%%%%%%%
\section{BWM Rate-Amplitude Constraints}

To synthesize the results from both our Earth- and pulsar-term analyses, we derive constraints on $\Lambda(>h_B)$, the annual rate of BWMs from any part of the sky with any polarization having amplitudes greater than $h_B$, assuming they occur as a Poisson process.  This quantity is readily relatable to astrophysical models of processes producing BWMs, i.e., Equation 15 of \citet{cj12}, Equation 21 of \citet{mcc14}, or Equation 17 of \cite{whc+14}. 

Toward this end, define the quantities,
\begin{eqnarray}
\tau_E(h_B,n)&=&\frac{\Delta t\Delta \Omega \Delta \psi}{4\pi^2}\sum_i^{N_t}\sum_j^{N_\Omega}\sum_k^{N_\psi}\Theta(h_B-n\sigma_{h,ijk}),\nonumber\\
\end{eqnarray}
and
\begin{eqnarray}
\tau_P(h_B,n)&=&\frac{\Delta t}{4\pi^2}\sum_K^{M}\sum_i^{N_t}\int\Theta\left(h_B-\frac{n\sigma_{h,p,K}}{B(\theta,\phi)}\right)d\psi d\Omega,\nonumber\\
\end{eqnarray}
where $\Delta t$, $\Delta \Omega$, and $\Delta \psi$ describe the grid spacing in the Earth-term search. The quantities $\tau_E$ and $\tau_P$ are the total time that the PTA had $n$-$\sigma$ sensitivity to a BWM of amplitude at least $h_B$ in the Earth-term and pulsar-term, respectively, weighted by the fraction of the total source-location and polarization angle space over which that sensitivity was achieved. Our definitions for $\tau_E$ and $\tau_P$ differ from nearly identical definitions in \citet{mcc14} by an overall factor of two. In that work, it was mistakenly assumed that only BWM polarization angles between $0$ and $\pi/2$ must be considered. We here correctly carry out a search that tests polarization angles between $0$ and $\pi$. 

With the definitions for $\tau_E$ and $\tau_P$ in place, we can derive constraints on $\Lambda(>h_B)$ from our Earth-term and pulsar-term analyses:
\begin{eqnarray}
\Lambda(>h_B) < -\frac{\ln{(1-Q)}}{\tau_{i}},
\end{eqnarray}
where $Q$ is the probability that at least one BWM occurring at rate $\Lambda(>h_B)$ encounters the PTA during the time $\tau_i$ and $i$ is a placeholder for either $E$ or $P$.

In Figure 7, we show our constraints on $\Lambda(>h_B)$ from our Earth- and pulsar-term analyses. We have set $Q=0.95$. We have set $n=3.52$ for both our Earth- and pulsar-term constraints. This number comes from requiring a false alarm probability of less than five-percent in an Earth-term search with $N_G=120$. Our estimate of $N_G$ is likely significantly larger than it needs to be. A detailed suite of simulations could give us a more realistic assessment of $N_G$ which would lead to more constraining upper limits, but for a fixed allowable false alarm probability, the amplitude a BWM must exceed to be inconsistent with the noise scales as the square root of the natural logarithm of $N_G$, so the improvement to the upper limit from this analysis would be very slight.

Since there are fewer effectively independent trials in a pulsar-term search, we should use a lower value of $n$ for a 95\% confidence upper limit. But, by treating the Earth- and pulsar-terms comparably, we are able to highlight an important fact about our data set. The near-total convergence of our Earth- and pulsar-term constraints for low values of $h_B$ indicates that our upper limits are almost entirely dominated by a single pulsar data set: J1713$+$0747. This was anticipated by \citet{mcc14} and holds true even with the more sophisticated noise modeling we have employed in this analysis.

\begin{figure}
\includegraphics[height=70mm]{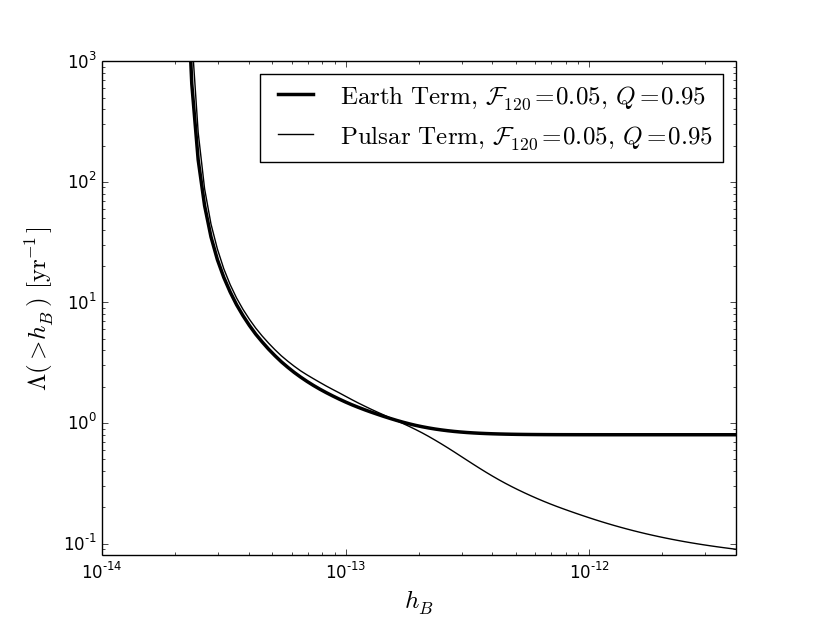}
\caption{95\% confidence Earth-term upper bound on the rate, $\Lambda(>h_B)$, of BWMs occurring with amplitudes at or above amplitudes $h_B$. The two curves come from our Earth-term and pulsar-term analyses. The pulsar-term probes lower-rate events at high amplitudes because individual pulsar terms contain causally independent information. The near total convergence of the Earth- and pulsar-term curves at low BWM amplitudes demonstrates that a single pulsar, J1713$+$0747, is dominating our sensitivity.}
\end{figure}

%%%%%%%%%%%%%%%%%%%%%%%%%%%%%%%%%%%%%%%%%%%%%%%%%%%%%%%%%%%%%%%%%%%%%%%%%%%
\section{Conclusion}

In this paper, we have conducted a search for BWMs in the first five years of NANOGrav data. We did not detect any BWMs. Based on our current understanding of SMBHBs, the most conventional anticipated source of bright BWMs, it is unsurprising that we did not. \citet{whc+14} predict that BWMs from SMBHB mergers exceeding amplitudes of $10^{-14}$ occur at a rate of just a few every $10^5$ yr. However, \citet{cj12} conclude that because of large uncertainties in things such as the inspiral rate of SMBHBs, the actual rate of BWMs is essentially unconstrained. Also, we stress that memory is a very general feature of bright GW events and large-amplitude BWMs may be produced by wholly unexpected phenomena occurring at unconstrained rates; ongoing searches for BWMs are crucial for utilizing the raw discovery potential of PTA observations \citep{cbv+14}. The methods developed in this paper are readily generalizable to future, more sensitive data sets and the accelerated search techniques we have developed for Earth-term analyses will greatly expedite future BWM searches.

Our Figure 7 and Figure 10 from \citet{whc+14} show constraints on the rate of BWMs from initial NANOGrav and PPTA data releases, respectively. The NANOGrav constraints and the PPTA constraints are quite similar. Both probe to BWM amplitudes of approximately $2\times 10^{-14}$; the PPTA upper limit extends to lower rates than the NANOGrav upper limit in part because the PPTA analyzed a slightly longer data span than we have analyzed here. Our Figure 6 and Figure 9 from \citet{whc+14} illustrate the sensitivity of NANOGrav and the PPTA to BWMs as it varies over the sky. Combined, these figures help support the science case of the IPTA. The area of the sky where NANOGrav's sensitivity to BWMs is worst is non-concentric with the area of the sky where the PPTA's sensitivity is worst. Joint analysis of data from NANOGrav and the PPTA will lead to more uniform sensitivity to BWMs over the whole sky and more constraining upper limits. Though the EPTA has not yet conducted a search for BWMs with their data, the inclusion of their data in the joint IPTA data set will likely play a similarly important role in improving the uniformity of the IPTA's BWM sensitivity. 

NANOGrav has made great strides in improving its BWM sensitivity since the collection of this initial data set. The pulsar timing backends at Arecibo and Green Bank, ASP and GASP, have been upgraded to the Puerto Rican Ultimate Pulsar Processing Instrument (PUPPI) and the Green Bank Ultimate Pulsar Processing Instrument \citep[GUPPI;][]{drd+08}, backends with exceptionally wide bandwidths that have reduced the rms timing errors on most pulsars being timed by a factor of two to three. Furthermore, NANOGrav is now regularly timing more than 40 pulsars rather than just 17 as in the first five years. Finally, just having a longer timing baseline on the pulsars we have analyzed in this paper is a great boon to our BWM sensitivity. All else being equal, sensitivity to BWMs scales approximately as $T^{-3/2}$, where $T$ is the span of the data set. If we then focus on SMBHBs above a minimum reduced mass (as in Figure 6), the volume of space in which we are sensitive to memory from their mergers grows as $T^{9/2}$; as the volume in which we are sensitive to billion solar mass mergers already encompasses parts of the Virgo cluster, this strong scaling of volume probed with time means that many more astrophysically interesting systems will enter our detection horizon with continued PTA observations.   NANOGrav is preparing a collection of nine years worth of data that will be a significantly more sensitive probe of BWMs than any PTA data set that has yet been analyzed. We plan to apply the techniques used in this paper to the nine-year NANOGrav data set to produce unprecedented constraints on BWMs.

%%%%%%%%%%%%%%%%%%%%%%%%%%%%%%%%%%%%%%%%%%%%%%%%%%%%%%%%%%%%%%%%%%%%%%%%%%%

\begin{acknowledgements}
We thank T. Loredo for helping us to appropriately treat extreme-value statistics and G. Hobbs for his maintenance of and continual improvements to \texttt{TEMPO2}. The work of Z.A., A.B., S.B.-S., S.J.C., S.C., B.C., N.J.C. J.M.C., P.B.D., T.D., J.A.E., N.G.-D., F.J., G.J., M.T.L., T.J.W.L., L.L., A.N.L., D.R.L., J.L., R.S.L., D.R.M., M.A.M., S.T.M., D.J.N., N.P., T.T.P., S.M.R., X.S., D.R.S., K.S., J.S., M.V., R.vH. and Y.W. was partially supported through the National Science Foundation (NSF) PIRE program award number 0968296. NANOGrav research at UBC is supported by an NSERC Discovery Grant and Discovery Accelerator Supplement and by the Canadian Institute for Advanced Research. D.R.M. acknowledges partial support through the New York Space Grant Consortium. J.A.E. and R.vH. acknowledge support by NASA through Einstein Fellowship grants PF4-150120 and PF3-140116, respectively. M.V. acknowledges support from the JPL RTD program. Portions of this research were carried out at the Jet Propulsion Laboratory, California Institute of Technology, under a contract with the National Aeronautics and Space Administration. Parts of the analysis in this work were carried out on the Nimrod cluster of S.M.R. Data for this project were collected using the facilities of the National Radio Astronomy Observatory and the Arecibo Observatory. The National Radio Astronomy Observatory is a facility of the NSF operated under cooperative agreement by Associated Universities, Inc. The Arecibo Observatory is operated by SRI International under a cooperative agreement with the NSF (AST-1100968), and in alliance with Ana G. M\'endez-Universidad Metropolitana and the Universities Space Research Association.  
\end{acknowledgements}

%%%%%%%%%%%%%%%%%%%%%%%%%%%%%%%%%%%%%%%%%%%%%%%%%%%%%%%%%%%%%%%%%%%%%%%%%%%

\appendix

\begin{longtable}{cclcc}
\tablewidth{0pc}
\tablecolumns{5}
\tablecaption{Table of Key Symbols}
\tablehead{
	\colhead{Symbol}&
	\colhead{}&
	\colhead{Description}&
	\colhead{}&
	\colhead{Dimensions}
	}

$\hat{~ ~}$ &$\cdots$& a ``hat" on any quantity indicates that the BWM amplitude used to calculate it is&$\cdots$&dimensionless\\&&the result of least-squares fitting && \\ 

$\tilde{~ ~}$ &$\cdots$& a ``tilde" on any quantity indicates that the BWM amplitude used to calculate it maximizes&$\cdots$&dimensionless\\&&the global likelihood (i.e., Equation 20) among trial amplitudes in a grid search   && \\ 

%$A$&$\cdots$& The amplitude of the power spectrum of a power-law&$\cdots$&time$^3$\\&&noise process influencing the timing residuals. && \\\\

%${\bf A}$ &$\cdots$& an operator mapping timing residuals to epoch-averaged residuals. &$\cdots$& dimensionless \\\\

$B(\theta,\phi)$ &$\cdots$& projection factor accounting for the geometric configuration of the Earth,&$\cdots$&dimensionless\\&& pulsar, and burst source and the burst polarization angle &&\\

${\bf C}_{\nu\nu',\tau\tau'}$ &$\cdots$& covariance from noise between a TOA from epoch $\tau$ and frequency&$\cdots$&time$^2$\\&&channel $\nu$ and a TOA from epoch $\tau'$ and frequency channel $\nu'$ && \\

%$\hat{D}$ &$\cdots$& $\cdots$ && \\

%$\hat{D}_{\rm max}$ &$\cdots$& $\cdots$ && \\

%$\hat{D}_{\rm thresh}$ && $\cdots$ && \\

%$\tilde{D}_G$ &$\cdots$& $\cdots$ && \\

$D(h_p,t_B)$ &$\cdots$& $2\ln{\Gamma}(h_p,t_B)$ &$\cdots$& dimensionless\\

$D_G$ &$\cdots$& $2\ln{\Gamma_G}$ &$\cdots$& dimensionless\\

$D_L$ &$\cdots$& luminosity distance from Earth to BWM source &$\cdots$& distance\\

$E$ &$\cdots$& EFAC, a constant multiplier on the rms perturbation attributed to radiometer noise &$\cdots$& dimensionless \\

$f_k$ &$\cdots$& the p.d.f. for the maximum of $k$ random numbers drawn from a $\chi^2$&$\cdots$&dimensionless\\&& distribution with one degree of freedom &&  \\

$F_k$ &$\cdots$& the c.d.f. associated with $f_k$ &$\cdots$& dimensionless \\

${\cal F}_k$ &$\cdots$& one minus the c.d.f. associated with $f_k$  &$\cdots$&dimensionless \\

$h_B$ &$\cdots$& amplitude of a BWM &$\cdots$& dimensionless\\

%$\hat{h}_B$ &$\cdots$& $\cdots$ && \\

$h_p$ &$\cdots$& projected BWM amplitude, $\pm h_BB(\theta,\phi)$ &$\cdots$& dimensionless\\

%${\cal I}$ &$\cdots$& inclination angle of a SMBHB just prior to merger && dimensionless\\

${\cal J}$ &$\cdots$& jitter-like covariance between TOAs from the same epoch but different frequency channels &$\cdots$& time \\

%${\cal L}({\bf R}|{\bf \delta p},{\bf \Xi})$ &$\cdots$& likelihood of residuals, ${\bf R}$, allowing for adjustments to the timing model\\&&parameters, ${\bf \delta p}$, and the noise model parameters, ${\bf \Xi}$ && dimensionless \\

%$M$ &$\cdots$& $\cdots$ && \\

%${\bf M}$ &$\cdots$& a timing model design matrix && mixed \\

%$N$ &$\cdots$& $\cdots$ && \\

%${\cal N}$ &$\cdots$& number of pulses integrated to form a TOA && dimensionless \\

%$N_h$ &$\cdots$& $\cdots$ && \\

%$N_p$ &$\cdots$& $\cdots$ && \\

%$N_t$ &$\cdots$& $\cdots$ && \\

%$N_\psi$ &$\cdots$& $\cdots$ && \\

%$N_\Omega$ &$\cdots$& $\cdots$ && \\

$N_I$ &$\cdots$& the effective number of trial BWM epochs tested in a single-pulsar BWM search &$\cdots$&dimensionless \\

$N_G$ &$\cdots$& the effective number of trials in the space of BWM epoch, sky position,&$\cdots$&dimensionless\\&&and polarization angle in a global Earth-term BWM search && \\

%$N_{\rm TOA}$ &$\cdots$& the total number of TOAs in a single pulsar timing data set && dimensionless \\

${\cal Q}$ &$\cdots$& EQUAD, the rms of a Gaussian noise process added in quadrature to radiometer noise &$\cdots$& time\\

%${\bf R}$ &$\cdots$& vector of timing residuals &$\cdots$& time \\

%${\bf R}_E$ &$\cdots$& vector of epoch-averaged timing residuals &$\cdots$& time \\\\

$t_0$ &$\cdots$& time at which BWM wavefront encounters the Earth &$\cdots$& time\\

$t_1$ &$\cdots$& time at which BWM wavefront is observed from Earth to encounter a pulsar &$\cdots$& time\\

$t_B$ &$\cdots$& epoch BWM is observed to occur, either $t_0$ or $t_1$&$\cdots$& time\\

%$t_{\nu,\tau}$ &$\cdots$& TOA for observing epoch $\tau$ and frequency channel $\nu$ && time\\

%$t_{\infty,\tau}$ &$\cdots$& infinite radio frequency TOA for observing epoch $\tau$ && time\\

%$t_{{\rm DM}_{\nu,\tau}}$ &$\cdots$& dispersive delay for observing epoch $\tau$ and frequency channel $\nu$ && time\\

%$t_{C_{\nu,\tau}}$ &$\cdots$& non-dispersive chromatic delays for observing epoch $\tau$ and frequency channel $\nu$ && time\\

${\bf U}$ &$\cdots$& the ``exploder" matrix that maps observation epochs to the full set&$\cdots$&dimensionless\\&&of TOAs, as discussed in \citet{abb+14} && \\

%$\gamma$ &$\cdots$& the spectral index of the power spectrum of a power-law\\&&noise process influencing the timing residuals && dimensionless \\

$\Gamma(h_p,t_B)$ &$\cdots$& likelihood of a pulsar's TOAs assuming a BWM of projected amplitude  $h_p$ occurred&$\cdots$&dimensionless\\&&at epoch $t_B$ divided by the likelihood of the TOAs assuming no BWM occurred &&  \\

%$\hat{\Gamma}$ &$\cdots$& $\cdots$ && \\

$\Gamma_G$ &$\cdots$& the global likelihood ratio, or product of pulsar-wise likelihood ratios (i.e., Equation 20) &$\cdots$&dimensionless\\&&for a trial BWM of fixed sky location, polarization, epoch, and amplitude&& \\

%${\bf \delta p}$ &$\cdots$& vector of modifications to a timing model && mixed \\

$\Delta t$ &$\cdots$& perturbation to pulse times of arrival &$\cdots$& time\\

%$\epsilon_{{\rm S/N}_{\nu,\tau}}$ &$\cdots$& timing perturbation from radiometer noise for observing epoch $\tau$ and frequency channel $\nu$ && time \\

%$\epsilon_{J_{\nu,\tau}}$ &$\cdots$& timing perturbation from pulse phase jitter for observing epoch $\tau$ and frequency channel $\nu$ && time \\

%$\epsilon_{{\rm DISS}_{\nu,\tau}}$ &$\cdots$& timing perturbation from diffractive interstellar scintillation\\&& for observing epoch $\tau$ and frequency channel $\nu$ && time \\

%$\epsilon_{\nu,\tau}$ &$\cdots$& the sum of all noise processes influencing the TOA from epoch $\tau$ and frequency channel $\nu$ && time \\

%$\theta$ &$\cdots$& angle between burst source direction and line of sight to pulsar && dimensionless \\

$\Lambda(>h_B)$ &$\cdots$& the rate that BWMs with amplitudes greater than $h_B$ encounter our PTA &$\cdots$& time$^{-1}$ \\

$\mu$ &$\cdots$& reduced mass of a SMBHB, $M_1M_2/(M_1+M_2)$ &$\cdots$& mass\\

%$\nu$ &$\cdots$& center frequency of a radio frequency channel && time$^{-1}$ \\

%${\bf \Xi}$ &$\cdots$& a vector of noise model parameters $A$, $\gamma$, ${\cal Q}$, ${\cal J}$, and $E$ && mixed \\

%$\rho_{J_{\nu,\nu'}}$ &$\cdots$& correlation coefficient of jitter noise between frequency channels $\nu$ and $\nu'$ && dimensionless \\

$\sigma_h$ &$\cdots$& uncertainty on the amplitude of a BWM&$\cdots$&dimensionless\\

$\sigma_{h,p}$ &$\cdots$& uncertainty on the projected amplitude of a BWM&$\cdots$& dimensionless \\

$\sigma_J$ &$\cdots$& rms timing perturbation from pulse phase jitter noise &$\cdots$& time \\

$\sigma_{\rm S/N}$ &$\cdots$& rms timing perturbation from radiometer noise &$\cdots$& time \\

%$\tau$ &$\cdots$& the epoch at which a TOA measurement is made && time \\

$\tau_E(h_B,n)$ &$\cdots$& total time that an Earth-term search could yield $n$-$\sigma_h$ detections of&$\cdots$&time\\&&a BWM with an amplitude greater than $h_B$ \\

$\tau_P(h_B,n)$ &$\cdots$& total time that a pulsar-term search could yield $n$-$\sigma_h$ detections of&$\cdots$& time\\&&a BWM with an amplitude greater than $h_B$ \\

%$\phi$ &$\cdots$& angle between the principle polarization vector of the burst and the projection of the  \\&& line of sight to the pulsar onto the plane normal to the burst propagation direction && dimensionless \\

%$\varphi$ &$\cdots$& phase of the pulse train of a pulsar as measured on Earth && dimensionless \\

$\chi^2(h_p,t_B)$ &$\cdots$& the square of the timing residuals when a BWM is included in the timing model&$\cdots$&dimensionless\\&&weighted by the inverse noise covariance matrix (see Equation 13) &&

%$\psi$ &$\cdots$& $\cdots$ &$\cdots$& dimensionless \\

%$\Omega$ &$\cdots$& $\cdots$ &$\cdots$& dimensionless \\

\end{longtable}

\tablecomments{For reference, descriptions of important or frequently used symbols.}

\bibliography{BWMLimit.bib}
\end{document}